\documentclass[showpacs,aps,graphicx,twocolumn]{revtex4}
\usepackage{mathrsfs}
\usepackage{amsfonts}
\usepackage{amssymb}
\usepackage{graphicx}
\usepackage{hyperref}
\usepackage{eufrak}
\usepackage[titletoc]{appendix}

\begin{document}

\title{Elementary gates of ternary quantum logic circuit}

\author{Yao-Min Di$^{1}$\footnote{Email address:
Yaomindi@sina.com} and Hai-Rui Wei$^{1,2}$}
\address{$^1$School of Physics $\&$ Electronic Engineering, Xuzhou Normal
University, Xuzhou 221116,  China \\
$^2$Department of Physics, Beijing Normal University, Beijing
100875, China}

\date{\today }

\begin{abstract}
The elementary gates are basic building blocks of quantum logic
circuit. They should be simple, efficient, and easy to implement. In
this article, we propose the ternary controlled X (TCX) gate or the
ternary controlled Z (TCZ) gate as the two-qutrit elementary gate,
which is universal when assisted by  arbitrary one-qutrit gates.
Based on Cartan decomposition, we give the one-qutrit elementary
gates. Also, we discuss the physical implementation of these
elementary gates  and show that they are feasible with current
technology. Then we investigate the synthesis of some important
ternary gates, such as the ternary SWAP gate, ternary Toffoli gates
and Muthukrishnan-Stroud gates. Finally we extend these elementary
gates to a more general case for qudit systems. This work provides a
unified description for the synthesis of both binary and
multi-valued quantum circuits.
\end{abstract}


\maketitle
 \flushbottom

\section{Introduction}

Quantum computer has attracted a great deal of attention due to its
potentialities to solve classical NP problems in polynomial time. In
quantum computing, the algorithms are commonly described by the
quantum circuit model \cite{1}. In 1995, Barenco \emph{et al.}
showed that any binary quantum logic circuit can be decomposed into
a sequence of one-qubit gates and CNOT gates \cite{2}. The process
of constructing quantum circuits by these elementary gates is called
synthesis by many authors. The complexity of quantum circuit can be
measured in terms of the number of elementary gates required.
Achieving gate arrays of less complexity is crucial as it reduces
not only the resource but also the errors.

Most approaches to quantum computing use two-level quantum systems
(qubits). Recent studies have indicated that there are some
advantages to expand quantum computer from qubits to multi-level
system (qudits). Three level quantum systems, so called qutrits, are
the simplest multi-valued systems. There have been many proposals to
use multi-level quantum systems to implement the quantum computation
and other quantum information processes \cite{3,4,5,6,7,8}. In
experiment, there have been reports on their applications in
simplifying quantum gates \cite{9}, simulating physical system with
spin greater than 1/2 \cite{10}. Multi-level quantum systems have
been realized in many ways in the field of optics \cite{11,12,13}.
In solid-state devices, the experimental demonstrations of full
quantum state tomography of the qutrit have been reported recently
\cite{14,15}. But multi-valued quantum logic synthesis is still a
new and immature research area. The crucial issue which gates is
chosen as the elementary gate set of multi-valued quantum circuit is
not well solved. The elementary gates are the basic blocks for
constructing quantum logic circuit, and they should be universal,
simple, effective and easy to implement.

A number of works have been done on multi-valued logic synthesis by
some authors. In 2000, Muthukrishnan and Stroud investigated the
synthesis of multi-valued quantum circuit \cite{4} and showed that
two two-qudit gates, which are called Muthukrishnan-Stroud gates
now, together with the one-qudit gates, are universal for quantum
computing. In 2002, Perkowski, Al-Rabadi, and Kerntopf proposed a
set of generalized ternary gates (GTG gate) \cite{16} based on a
ternary condition gate and ternary shift gates \cite{17}. During
2005 to 2006, based on cosine-sine decomposition of matrix
\cite{18}, the synthesis of ternary and more general multi-valued
quantum logic circuits was investigated by Khan and Perkowski in
Refs.\cite{19} and \cite{20}, respectively. The multi-valued quantum
circuit can be synthesized in terms of quantum multiplexers and
uniformly controlled rotations. But these components themselves have
a complicated structure and their synthesis needs study further.

On the other hand, two Brylinskies \cite{21} proved that any
two-qudit gate that creates entanglement without ancillas can act as
a universal gate for quantum computation, when assisted by arbitrary
one-qudit gates. That is to say, ``almost every'' two-qudit gate is
universal when assisted by one-qudit gates. But not all these gates
are suitable to be chosen as the two-qudit elementary gate of the
quantum multi-valued circuit. Just as the binary quantum circuit, we
usually choose CNOT gate or controlled-Z gate as the two-qubit
elementary gate although ``almost every'' two-qubit gate is
universal. Two Brylinskies' proof relies on a long argument using
advanced mathematics. No any specific gate is proposed as the
two-qudit elementary gate.  Alber investigated the purification of
bipartite high dimension quantum states with hermitian generalized
XOR gate (GXOR gate)\emph{et al.} \cite{22}. As a universal
bipartite gate for ternary quantum computing, a protocol of the
physical implementation of the GXOR gate on ions in a trap was
presented by Klimov \emph{et al.} in  \cite{23}. Wang \emph{et al.}
discussed the entanglement power of operators in qudit systems. They
proposed to choose the SUM gate, which is called Feynman gate in
computer community, as the elementary bipartite gate for qudit
quantum computing \cite{24}. But little work has done based on these
gates for the specific synthesis of multi-valued quantum circuit.

In this article, we focus on the investigation of elementary gates
of ternary quantum logic circuits. We propose the ternary controlled
X (TCX) gate or the ternary controlled Z (TCZ) gate as the
two-qutrit elementary gate. Based on the Cartan decomposition
\cite{25}, the one-qutrit elementary gates are also given. The
elementary gates proposed here are essentially binary and which can
be implemented with current technology. Also, we expand these
elementary gates to a more general case of qudit systems. So many
results in binary quantum logic circuits can be generalized to a
multi-valued case. They can be used as a unified measure of
complexity for various quantum logic circuits.

This article is organized as follows. In Sec.
\uppercase\expandafter{\romannumeral2}, we investigate two-qutrit
elementary gate for ternary quantum logic circuits and propose the
TCX gate or the TCZ gate as the elementary gate. In Sec.
\uppercase\expandafter{\romannumeral3}, based on Cartan
decomposition we discuss the set of one-qutrit elementary gates and
the synthesis of generic one-qutrit gates. The physical
implementations of these elementary gates are studied in Sec.
\uppercase\expandafter{\romannumeral4}. And the synthesis of some
important ternary quantum gates, such as the ternary SWAP gate,
ternary Toffoli gates and Muthukrishnan-Stroud gates, is given in
Sec. \uppercase\expandafter{\romannumeral5}. In Sec.
\uppercase\expandafter{\romannumeral6}, we extend our study to a
general case for qudit systems. Finally a brief conclusion and
future work is given in Sec. \uppercase\expandafter{\romannumeral7}.

\section{Two-qutrit elementary gates } 

In one qutrit case, there are three X quantum gates given by the
matrices
\begin{eqnarray}                                      \label{1}
X^{(01)}=\left(\begin{array}{ccc}
0&1&0\\
1&0&0\\
0&0&1\\
\end{array}\right), \nonumber \\
X^{(02)}=\left(\begin{array}{ccc}
0&0&1\\
0&1&0\\
1&0&0\\
\end{array}\right), \\
X^{(12)}=\left(\begin{array}{ccc}
1&0&0\\
0&0&1\\
0&1&0\\
\end{array}\right). \nonumber
\end{eqnarray}
Similarly, for every single qubit gate A, we can simply extend it to
a set of ternary gate, $A^{(ij)}$. The ternary extension of Hadamard
gate can be expressed as
\begin{eqnarray}                                      \label{2}
H^{(01)}=\frac{1}{\sqrt{2}}\left(\begin{array}{ccc}
1&1&0\\
1&-1&0\\
0&0& \sqrt{2}\\
\end{array}\right), \nonumber \\
H^{(02)}=\frac{1}{\sqrt{2}}\left(\begin{array}{ccc}
1&0&1\\
0&\sqrt{2}&0\\
1&0&-1\\
\end{array}\right), \\
H^{(12)}=\frac{1}{\sqrt{2}}\left(\begin{array}{ccc}
\sqrt{2}&0&0\\
0&1&1\\
0&1&-1\\
\end{array}\right). \nonumber
\end{eqnarray}
The extension of Z gate is slightly different, we denote them as
\begin{eqnarray}                                    \label{3}
&& Z^{[0]}=diag \{-1,I_{2}\},
 Z^{[1]}=diag\{1,-1,1\}, \nonumber \\
&&Z^{[2]}=diag \{I_{2},-1\}.
\end{eqnarray}

\begin{figure}[!h]
\begin{center}
\includegraphics[width=2.5 cm,angle=0]{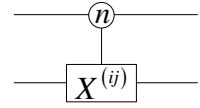}
\caption{Ternary controlled-X gate.}  \label{Fig1}                                     
\end{center}
\end{figure}

\begin{center}
\begin{figure}[!h]
\begin{center}
\includegraphics[width=6.5 cm,angle=0]{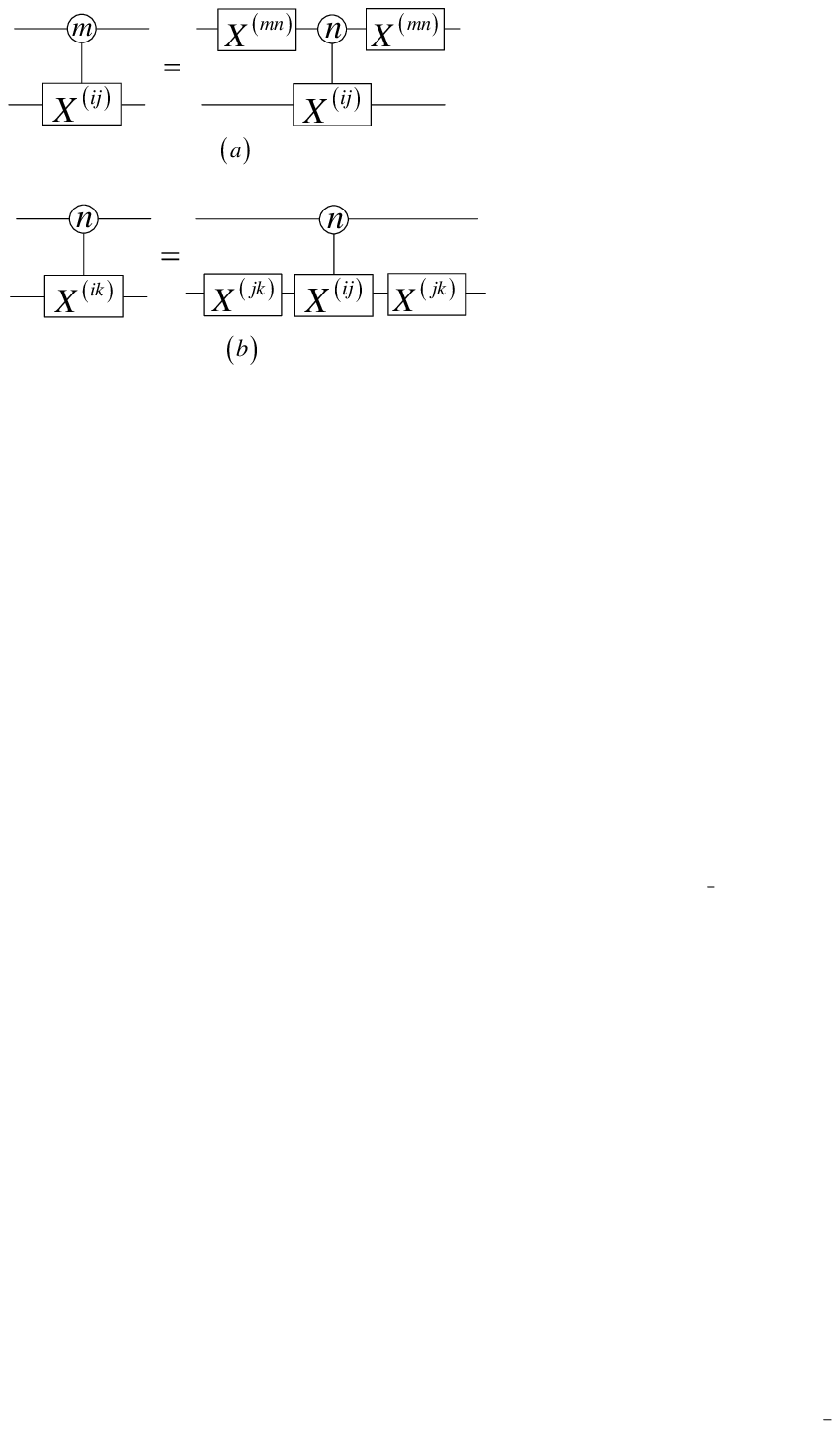}
\caption{Transformation among different TCX gates.
 (a) Transformation of control mode. (b) Transformation of target operations.}    \label{Fig2}                          
\end{center}
\end{figure}
\end{center}

\begin{figure}[!h]
\begin{center}
\includegraphics[width=7.0 cm,angle=0]{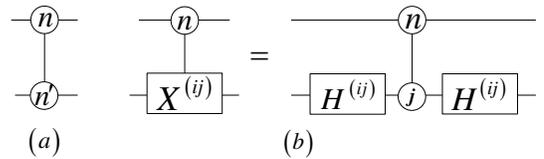}
\caption{TCZ gates and transformation between TCX and TCZ gates.}   \label{Fig3}                      
\end{center}
\end{figure}

The TCX gate is a two-qutrit gate which is defined that the gate
implements the $X^{(ij)}$ operation on the target qutrit iff the
control qutrit is in the states $|n\rangle$, $(n\in\{ 0, 1, 2\})$.
The circuit representation for the TCX gate is shown in Fig.
\ref{Fig1} in which the line with a circle represents the control
qutrit, while that with a square represents the target qutrit. There
are nine different forms for the TCX gate and they can be easily
transferred one another as shown in Fig. \ref{Fig2}. The TCZ gate is
defined that the gate implements the $Z^{[n']}$ operation on the
target qutrit iff the control qutrit is in the states $|n\rangle$.
Similar to the binary controlled Z gate, the control qutrit and
target qutrit of TCZ gate are exchangeable. It has nine different
forms, which also can be transferred one another by using ternary X
gates. The symbol of TCZ gate and its transformation relation with
TCX gate are shown in Fig. \ref{Fig3}.

\begin{figure}[!h]
\begin{center}
\includegraphics[width=8.0 cm,angle=0]{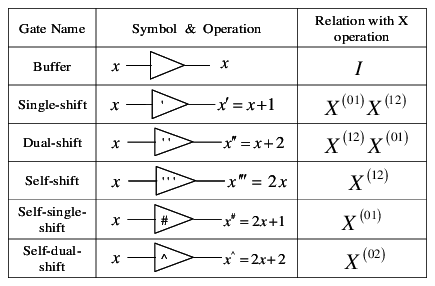}
\caption{Ternary shift gates.}   \label{Fig4}                                                             
\end{center}
\end{figure}

\begin{figure}[!h]
\begin{center}
\includegraphics[width=4.0 cm,angle=0]{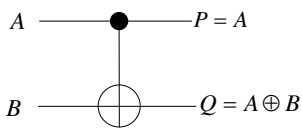}
\caption{Ternary Feyman gate.}\label{Fig5}                                                                
\end{center}
\end{figure}

\begin{figure}[!h]
\begin{center}
\includegraphics[width=6.0 cm,angle=0]{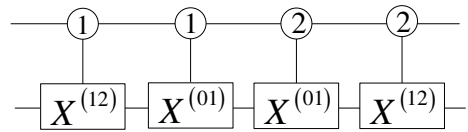}
\caption{Synthesis of ternary Feynman gate.} \label{Fig6}                                      
\end{center}
\end{figure}

\begin{figure}[!h]
\begin{center}
\includegraphics[width=4.5 cm,angle=0]{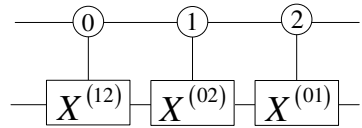}
\caption{Synthesis of GXOR gate.}  \label{Fig7}                                          
\end{center}
\end{figure}

The ternary shift gates proposed in Ref. \cite{17} are basic
one-qutrit gates. Operations, symbols and the relations with X
operations are listed in Fig. \ref{Fig4}. Likewise, the two-qutrit
controlled shift gates can be defined. Two-qutrit Feynman gate which
is called SUM gate in \cite{24} is shown in Fig. \ref{Fig5}. Here
$A$ is the controlling input and $B$ is the controlled input. The
output in control qutrit equals to the input $A$, and the output in
target qutrit is the sum of $A$ and $B$ modulo 3. The synthesis of
the ternary Feynman gate base on the TCX gates is shown in Fig.
\ref{Fig6}. GXOR gate is similar to the Feynman gate, and the
difference is that the output is the difference of $A$ and $B$
modulo 3. Its synthesis is shown in Fig. \ref{Fig7}. The GTG gates
are the combinations of the controlled shift gates and they are
mainly used  to investigate the synthesis of permutation quantum
gates by some groups \cite{16,17}. The permutation gate is a gate
which unitary matrices have only one 1 in every column and the
remaining elements 0.

The X operation is more elementary than the shift operation. From
the point of view of group theory, the six shift operations
constitutes a permutation group $\mathrm{S}_{3}$, and the $X^{(01)}$
and $X^{(12)}$ operations are the generators of the group. The TCX
gate is an elementary counterpart of the binary CNOT gate. It is
simple and easy to implement. While the Feynman gate, GXOR gate and
GTG gates have complex structures themselves. So we propose the TCX
gate as the two-qutrit elementary gate for qutrit-based quantum
computation. Of course, the choice is not unique, and the TCZ gate
also can be chosen as the two-qutrit elementary gate.

\section{One-qutrit elementary gates} 

The complexity of binary quantum logic gate is usually measured by
the numbers of CNOT gate and one-qubit $R_y$, $R_z$ gates. We call
the $R_y$, $R_z$ gates as one-qubit elementary gates. Suppose $M$ is
the matrix of a one-qutrit gate. Using the Cartan decomposition of
Lie group, it can be expressed \cite{26} as
\begin{eqnarray}                                                           \label{4}
M&=&e^{i\alpha} R_{y}^{(01)}(\beta) R_{y}^{(02)}(\gamma)
R_{y}^{(01)}(\delta) R_{z}^{(01)}(\theta)
R_{z}^{(02)}(\varphi)\nonumber \\
&& R_{y}^{(01)}(\beta') R_{y}^{(02)}(\gamma') R_{y}^{(01)}(\delta'),
\end{eqnarray}
where $\alpha, \beta, \gamma$, etc. are all real numbers.  Here
$R_\alpha^{(jk)}(\theta)=\exp(-i\theta\sigma_\alpha^{(jk)}/2)$ for
$j<k$ and $\alpha\in \{x,y,z\}$. $\sigma_x^{(jk)}=|j\rangle\langle
k|+|k\rangle\langle j|$, $\sigma_y^{(jk)}=-i|j\rangle\langle
k|+i|k\rangle\langle j|$, and $\sigma_z^{(jk)}=|j\rangle\langle
j|-|k\rangle\langle k|$. The four basic one-qutrit gates,
$R_y^{(01)}$, $R_z^{(01)}$, $R_y^{(02)}$, $R_z^{(02)}$ constitutes a
set of one-qutrit elementary gates.

The Cartan decomposition of one-qutrit gates is not unique, so the
choice of one-qutrit elementary gates is not unique too. From one
kind of Cartan decomposition presented in Appendix A, we get another
product expression of a single qutrit gate that
\begin{eqnarray}                                                         \label{5}
M=e^{i\varphi}M_1^{(jk)}M^{(j'k')}M_2^{(jk)}.
\end{eqnarray}
Here $M^{(jk)}$ is a special unitary transformation in 2-dimensional
subspace $\mathscr{H}_{jk}$, and it can be factored further by the
Euler decomposition. The Euler decomposition usually has two modes:
ZYZ decomposition and XYX decomposition. So the set of one-qutrit
elementary gates has two pairs of basic gates in subspaces
$\mathscr{H}_{jk}$ and  $\mathscr{H}_{j'k'}$ respectively. We can
take $R_y^{(jk)}$, $R_z^{(jk)}$, $R_y^{(j'k')}$, $R_z^{(j'k')}$ or
$R_x^{(jk)}$, $R_y^{(jk)}$, $R_x^{(j'k')}$, $R_y^{(j'k')}$ as
one-qutrit elementary gates. The set of one-qutrit elementary gate
given in Ref. \cite{26} is one of them. The synthesis of generic
one-qutrit gates is given by Eq. (\ref{4}) or Eq. (\ref{5}).

\section{Physical implementation of ternary elementary gates}  

As we know, there are not many proposals on the physical
implementation of ternary gates. In Ref. \cite{4}, a scheme for the
implementation of Muthukrishnan-Stroud gates based on the linear ion
trap is given. Also based on the ion trap, a scheme for the GXOR
gate is given by Klimov \emph{et al.} in Ref. \cite{23}. But these
schemes and the gates themselves are rather complicated,  and no
experimental investigations on them have been reported yet. However,
in the last decade, there has been tremendous progress in the
experimental development of binary quantum computing, and the
problem of constructing a CNOT gate has been addressed from various
perspectives and for different physical systems
\cite{27,28,29,30,31,32,33,34,35,36,37,38,39}. The elementary gates
proposed here can be implemented by existing technique.

\begin{figure}[!h]
\begin{center}
\includegraphics[width=4.5 cm,angle=0]{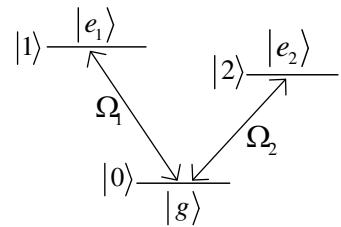}
\caption{V-type three level quantum system.}   \label{Fig8}                                             
\end{center}
\end{figure}

Assume we have a V-type three-level quantum system shown in Fig.
\ref{Fig8}, which constitutes a qutrit and the two levels of the
system $|0\rangle$ and $|1\rangle$ forms a qubit. Two laser beams
$\Omega_{1}$ and $\Omega_{2}$  are applied to the ion to manipulate
$|0\rangle\leftrightarrow|1\rangle$
  and $|0\rangle\leftrightarrow|2\rangle$ transition, respectively. If a two-qubit CNOT gate is realized
in such systems, one of TCX gate is naturally obtained, and the
eight other form TCX gates can be obtained by the transformation
shown in Fig. \ref{Fig2}. The single qutrit gates are implemented by
Rabi oscillations between the qutrit levels. Applying the laser
pulses in $\Omega_{1}$ and $\Omega_{2}$  and choosing suitable
phases, this allows us to perform $R_{x}^{(01)}$, $R_{y}^{(01)}$ and
$R_{x}^{(02)}$, $R_{y}^{(02)}$ gates respectively \cite{1,40}. So a
set of one-qutrit elementary gates is obtained, and any one-qutrit
gate can be implemented according to Eq. (\ref{4}) or Eq.
(\ref{A8}). There are other two types of quantum system,
$\Lambda$-type and cascade type. We can use $R_{x}^{(01)}$,
$R_{y}^{(01)}$, $R_{x}^{(12)}$, $R_{y}^{(12)}$ or $R_{y}^{(02)}$,
$R_{z}^{(02)}$, $R_{y}^{(12)}$, $R_{z}^{(12)}$ as one-qutrit
elementary gates to meet the requirement of manipulating quantum
states in these types of quantum system.

It is not too difficult to find such a quantum system. Early in
2003, the Innsbruck group implemented the complete Cirac-Zoller
protocol \cite{27} of CONT gate with two calcium ions
($\mathrm{Ca^{+}}$) in a trap \cite{29}. The energy level scheme of
$\mathrm{Ca^{+}}$ is given in Ref. \cite{31}. The original qubit
information is encoded in ground state $S_{1/2}$ and metastable
$D_{5/2}$ state. The $D_{5/2}$ state has a lifetime $\tau \backsimeq
1.16$ s. There is another metastable $D_{3/2}$ state in
$\mathrm{Ca^{+}}$. Its lifetime, which is measured recently by
Kreuter et al. \cite{41}, is about the same as that of the $D_{5/2}$
state. The three levels of $\mathrm{Ca^{+}}$, one ground state and
two metastable states, may constitute a qutrit candidate. The CNOT
gate was implemented by Schmidt-Kaler \emph{et al.} \cite{29} forms
naturally a TCX gate. Two laser pulses are used to manipulate the
$S_{1/2}\leftrightarrow D_{5/2}$ quadruple transition near 729 nm
and the $S_{1/2}\leftrightarrow D_{3/2}$ transition near 732 nm,
respectively. Rabi oscillations between these levels can implement
the one-qutrit elementary gates $R_x^{(01)}$, $R_y^{(01)}$ and
$R_x^{(02)}$, $R_y^{(02)}$.

The superconducting quantum information processing devices are
typically operated as qubit by restricting it to the two lowest
energy eigenstates. By relaxing this restriction, we can operate it
as a qutrit or qudit. As mentioned in introduction, the experimental
demonstrations of the tomography of a transmon-type superconducting
qutrit and a superconducting phase qutrit have been reported in Ref.
\cite{14} and Ref. \cite{15}, respectively. It means that to prepare
arbitrary one-qutrit state and read out with high-fidelity on these
systems has been implemented. So the one-qutrit gates can be
implemented on the systems by the method described here.
Construction of a robust CNOT gate on superconducting qubits has
been extensively investigated both in theory and experiment
\cite{33,34,35,36,37,38}. So the condition to implement elementary
gates of ternary quantum logic circuit has come to maturity on these
superconducting qutrits.

\section{Synthesis of some important ternary quantum gates}

Since the X operations given in Eq. (\ref{1}) and the four
one-qutrit elementary gates all only act on two levels in a qutrit,
many results in binary quantum logic circuit can be generalized to
ternary quantum logic circuits. The synthesis of binary quantum
circuit has been extensively investigated by many groups
\cite{1,2,42,43,44,45,46,47}, and it is rather mature now. The
ternary SWAP gate interchanges the states of two qutrits acted by
the gate. It is a generalization of binary SWAP gate and can be
decomposed into three binary SWAP gates, i.e.,
\begin{eqnarray}                                               \label{6}
W=W^{(01)}\cdot W^{(02)}\cdot W^{(12)}.
\end{eqnarray}
Here the $W^{(ij)}$ can be called as a conditional SWAP gate, and
each of them is synthesized by three TCX gates. So the ternary SWAP
gate is synthesized by nine TCX gates, as shown in Fig. \ref{Fig9}.
Likewise, the ternary root SWAP gate can be decomposed into three
binary root SWAP gates.

The ternary Toffoli gate  has many forms. Here we define an
elementary ternary Toffoli gate that two control qutrits are
unaffected by the action of the gate, and the target qutrit is acted
by the $X^{(ij)}$ operation iff the two control qutrits are in the
states $|n\rangle$, $|n'\rangle$ respectively. By the result of
quantum synthesis for the binary Toffoli gate \cite{1}, the
synthesis of the elementary ternary Toffoli gate can be obtained and
is illustrated in Fig. \ref{Fig10}. It needs six TCX gates and ten
single qutrit gates, which are the simple extension of single qubit
gates $H$, $T$, $T^{\dagger}$ and $S$.

\begin{widetext}
\begin{center}
\begin{figure}[!h]
\includegraphics[width=13.0 cm,angle=0]{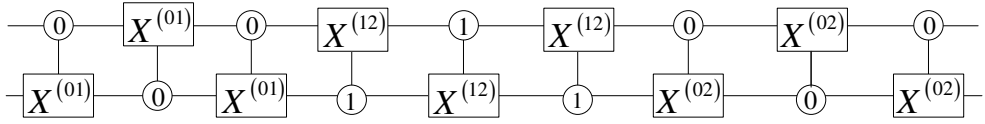}
\caption{Synthesis of ternary SWAP gate.} \label{Fig9}                                        
\end{figure}
\end{center}

\begin{figure}[!h]
\begin{center}
\includegraphics[width=16.0 cm,angle=0]{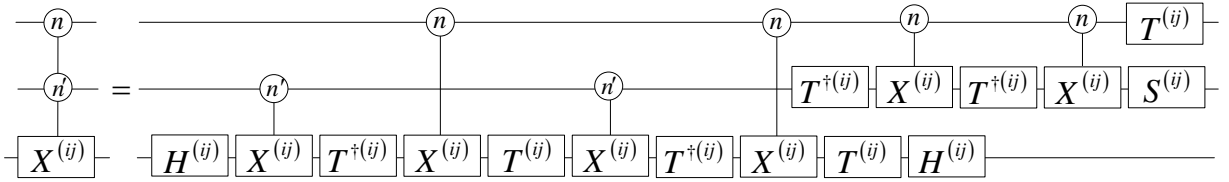}
\caption{Synthesis of ternary elementary Toffoli gate.}  \label{Fig10}                                  
\end{center}
\end{figure}

\begin{figure}[!h]
\begin{center}
\includegraphics[width=17.5 cm,angle=0]{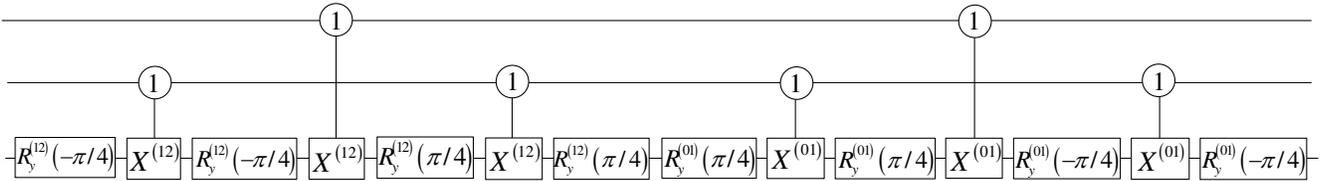}
\caption{Synthesis of a typical ternary Toffoli gate.} \label{Fig11}                                  
\end{center}
\end{figure}
\end{widetext}

A typical ternary Toffoli gate is that the two control qutrits
remain no change, and the output of target qutrit is $C\oplus1$ iff
both two qutrits are in the state $|1\rangle$, where $C$ is the
input of the target qutrit. Yang \emph{et al.} defined a generalized
ternary Toffoli gate for multiple qutrit systems in Ref. \cite{48},
and it is just this kind of ternary Toffoli gate in the three
qutrits case. We give its synthesis shown in Fig. \ref{Fig11}, which
needs six TCX gates and eight one-qutrit elementary gates.

The two Muthukrishnan-Stroud gates are denoted by
$\Gamma_{2}(\mathbb{Z})$ and $\Gamma_{2}(\Phi)$ respectively. They
are two-qutrit controlled gates in which  if the control qutrit is
set to $|2\rangle$, then the operation $\mathbb{Z}$ or $\Phi$ is
applied to the target qutrit respectively.  $\mathbb{Z}$ is a family
of one-qutrit gates which transfer a definite single qutrit state to
the state $|2\rangle$, that is,
\begin{eqnarray}                                                                         \label{7}
\mathbb{Z}(c_{0},c_{1},c_{2}):c_{0}|0\rangle+c_{1}|1\rangle+c_{2}|2\rangle\mapsto|2\rangle.
\end{eqnarray}
It does not determine the transform uniquely. Assume
$c_{0}=\cos\theta_{1}e^{i\varphi_{0}}$,
$c_{1}=\sin\theta_{1}\cos\theta_{2}e^{i\varphi_{1}}$, and
$c_{2}=\sin\theta_{1}\sin\theta_{2}$, one of expression of the
operation can be written as
\begin{eqnarray}                                                          \label{8}
\mathbb{Z}=PQ&=& \left(\begin{array}{ccc}
\sin\theta_{1}&0&-\cos\theta_{1}e^{i\varphi_{0}}\\
0&1&0\\
\cos\theta_{1}e^{-i\varphi_{0}}&0&\sin\theta_{1}\\
\end{array}\right)\times \nonumber \\&&
 \left(\begin{array}{ccc}
1&0&0\\
0&\sin\theta_{2}&-\cos\theta_{2}e^{i\varphi_{1}}\\
0&\cos\theta_{2}e^{-i\varphi_{1}}&\sin\theta_{2}\\
\end{array}\right).
\end{eqnarray}
The synthesis of  $\Gamma_{2}(\mathbb{Z})$ based on TCZ gate is
shown in Fig. 12. Here
\begin{eqnarray}                                                                                     \label{9}
V_{1}&=&\left(\begin{array}{ccc}
1&0&0\\
0&\cos(\frac{\pi}{4}-\frac{\theta_{2}}{2})&-\sin(\frac{\pi}{4}-\frac{\theta_{2}}{2})e^{i\varphi_{1}}\\
0&\sin(\frac{\pi}{4}-\frac{\theta_{2}}{2})e^{-i\varphi_{1}}&\cos(\frac{\pi}{4}-\frac{\theta_{2}}{2})\\
\end{array}\right)\nonumber \\
&=&R_{z}^{(12)}(-\varphi_{1}) R_{y}^{(12)}(\frac{\pi}{2}-\theta_{2})
R_{z}^{(12)}(\varphi_{1}),
\end{eqnarray}
\begin{eqnarray}                                                                                       \label{10}
V_{2}&=&\left(\begin{array}{ccc}
\cos(\frac{\pi}{4}-\frac{\theta_{1}}{2})&0&-\sin(\frac{\pi}{4}-\frac{\theta_{1}}{2})e^{i\varphi_{0}}\\
0&1&0\\
\sin(\frac{\pi}{4}-\frac{\theta_{1}}{2})e^{-i\varphi_{0}}&0&\cos(\frac{\pi}{4}-\frac{\theta_{1}}{2})\\
\end{array}\right) \nonumber \\
&=&R_{z}^{(02)}(-\varphi_{0}) R_{y}^{(02)}(\frac{\pi}{2}-\theta_{1})
R_{z}^{(02)}(\varphi_{0}).
\end{eqnarray}

The $\Phi$ is a single qutrit phase gate which advances the phase of
$|2\rangle$ without affecting $|0\rangle$ and $|1\rangle$ states in
the qutrit. $\Gamma_{2}(\Phi)=diag\{I_8,e^{i\varphi}\}$ and its
synthesis is shown in Fig. \ref{Fig13}.

\begin{figure}[!h]
\begin{center}
\includegraphics[width=7.0 cm,angle=0]{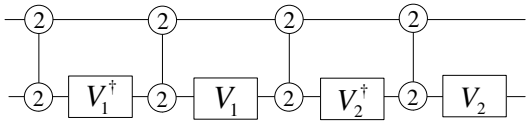}
\caption{Synthesis of Muthukrishnan-Stroud gate $\Gamma_{2}(\mathbb{Z})$.}  \label{Fig12}                
\end{center}
\end{figure}

\begin{figure}[!h]
\begin{center}
\includegraphics[width=8.5 cm,angle=0]{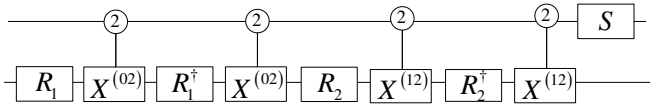}
\caption{Synthesis of Muthukrishnan-Stroud gate $\Gamma_{2}(\Phi)$,
where $S=diag\{I_2,e^{i\varphi/3}\}$, $R_1=R_z^{(02)}(\varphi/3)$, $R_1=R_z^{(12)}(\varphi/3)$.}  \label{Fig13}     
\end{center}
\end{figure}

Based on the CSD, Khan and Perkowski investigate the structure of
ternary quantum circuit. An arbitrary n qutrit gate can be
synthesized with four multiplexers acting on $n-1$ qutrits and three
$(n-1)$-fold uniformly controlled rotations. The syntheses of these
multiplexers and uniformly controlled rotations are much more
complicated than that of Muthukrishnan-Stroud gate. We will
investigate them in another article.

\section{General multi-valued case}

In this section, we  generalize the elementary gates to a general
multi-valued quantum logic circuit case. We first extend the single
qubit X gate to $d$-dimensional quantum systems (qudits) similar to
the ternary case. The single qudit gate $X^{(jk)}$ is a gate which
acts the X operation in two-dimensional subspace $\mathscr{H}_{jk}$
of $d$-dimensional Hilbert space. Similarly the definition of TCX
gate can be naturally generalized to the qudit case. The generalized
controlled X (GCX) gate is the two-qudit gate it implements the
$X^{(ij)}$ operation on the target qudit iff the control qudit is in
the states $|n\rangle$ ($n\in \{0, 1, \cdots, d-1\}$). Likewise, we
can define the generalized controlled Z (GCZ) gate. The GCX gate or
GCZ gate can be chosen as the two-qudit elementary gate, which is
universal for qudit quantum computing, when it is assisted by
arbitrary one-qudit gates. The GCX gate has $d$ kinds of control
mode and $\frac{1}{2}d(d-1)$ different $X^{(ij)}$ operations. They
can be transferred one another by the similar mode shown in Fig.
\ref{Fig2}. This holds true for the GCZ gate too.

The matrices of $d$-dimensional one-qudit gates are the elements of
$d$-dimensional unitary group. From successive Cartan decompositions
of  $\mathrm{U}(d)$ group and Euler decompositions, we can show that
the set of one-qudit elementary gates has $d-1$ pairs of basic gates
acting on $d-1$ different 2-dimensional subspaces. So  manipulating
a qudit completely needs at least $d-1$ driving fields. The choice
of $d-1$ pairs of basic gates is not unique. They are universal if
only the corresponding driving fields can connect the $d$ levels of
the qudit together. Like qutrit case, the pair of basic gates also
has two modes: $R_y$, $R_z$ and $R_x$, $R_y$  modes. The Cartan
decomposition of a one-qudit gate has given in appendix B.

\section{Conclusion and future work}

We have investigated the elementary gates for ternary quantum logic
circuits. We propose the TCX or the TCZ gates as a two-qutrit
elementary gate, with which arbitrary ternary quantum circuits can
be synthesized when they are assisted by single qutrit gates. Based
on the Cartan decomposition,   the one-qutrit elementary gates are
also investigated. They have two pairs of basic gates and two modes:
$R_y^{(jk)}$, $R_z^{(jk)}$, $R_y^{(j'k')}$, $R_z^{(j'k')}$, or
$R_x^{(jk)}$, $R_y^{(jk)}$, $R_x^{(j'k')}$, $R_y^{(j'k')}$. Then we
have discussed the implementation scheme for the ternary elementary
gates and have investigated the synthesis of some important ternary
gates. The elementary gates proposed here are simple, efficient and
can be implemented with current technology. Moreover, these
elementary gates can be easily extended to a more general qudit
case, so they constitute unified elements to synthesize quantum
logic circuits, whatever they are qubit, qutrit, qudit or hybrid
circuits. We can use them as a unified measure for the complexity of
various quantum circuits.

The multi-valued and hybrid quantum computing is a new and exciting
research area in which there is plenty of work to do. Moreover, to
choose suitable quantum system, such as trapped ions,
superconducting qutrits or qudits and quantum dots, to investigate
the physical implementation of multi-valued quantum logic gates and
to undertake the experimental work is crucial for the development of
multi-valued quantum information science.

\section*{ACKNOWLEDGEMENTS}

We would like to thank Dr. Fu-Guo Deng for his carefully reading the
manuscript, improving English, and helpful discussion. The work was
supported by the Project of Natural Science Foundation of Jiangsu
Education Bureau, China (Grant No. 09KJB140010).

\appendix

\section{CARTAN DECOMPOSITON OF ONE-QUTRIT GATES}
The Cartan decomposition of a Lie group depends on the decomposition
of its Lie algebras \cite{25}. Let $\mathfrak{g}$ be a semisimple
Lie algebra and there is the decomposition relations
\setcounter{equation}{0}
\begin{equation}                                    \label{A1}
\mathfrak{g}=\mathfrak{l}\oplus \mathfrak{p},
\end{equation}
where $\mathfrak{l}$ and $\mathfrak{p}$ satisfy the commutation
relations
\begin{eqnarray}                                                        \label{A2}
[\mathfrak{l},\mathfrak{l}]\subseteq \mathfrak{l},
[\mathfrak{l},\mathfrak{p}]\subseteq \mathfrak{p},
[\mathfrak{p},\mathfrak{p}]\subseteq \mathfrak{l},
\end{eqnarray}
we said the decomposition is the Cartan decomposition of Lie algebra
$\mathfrak{g}$. The $\mathfrak{l}$ is closed under the Lie bracket,
so it is a Lie subalgebra of $\mathfrak{g}$, and that
$\mathfrak{p}=\mathfrak{l}^{\bot}$. A maximal Abelian subalgebra
$\mathfrak{a}$ contained in $\mathfrak{p}$ is called a Cartan
subalgebra. Then using the relation between Lie group and Lie
algebra, every element $M$ of Lie group $G$ can be decomposed as
\begin{eqnarray}                                              \label{A3}
M=K_{1}AK_{2},
\end{eqnarray}
 where  $G=e^\mathfrak{g}$, $K_{1}, K_{2}\in
e^\mathfrak{l}$ and $A\in e^\mathfrak{a}$.

The one-qutrit gates form a $3$-dimensional unitary group
$\mathrm{U}(3)$. We have 8 independent ternary Pauli's matrices:
three $\sigma_{x}^{(ij)}$ matrices, three $\sigma_{y}^{(ij)}$
matrices, and the two independent $\sigma_{z}^{(ij)}$ matrices in
the three of them. Multiplying these 8 independent Pauli's matrices
by $i$, we get the basis vectors of Lie algebra $\mathrm{su}(3)$
which we called the qusi-spin basis. Together with $3\times3$
identity matrix multiplied by $i$, they constitute the basis vectors
of Lie algebra $\mathrm{u}(3)$. Take a \textbf{AIII} type Cartan
decomposition \cite{25} of $\mathrm{u}(3)$, that is
\begin{eqnarray}                                                                                     \label{A4}
\mathrm{u}(3)=\mathrm{s}(\mathrm{u}(2) \oplus \mathrm{u}(1))\oplus
\mathrm{s}(\mathrm{u}(2) \oplus \mathrm{u}(1))^{\bot}.
\end{eqnarray}
Lie subalgebra $\mathrm{s}(\mathrm{u}(2) \oplus \mathrm{u}(1))$
consists of subagebra $\mathrm{su}(2)$ and a complex basis
$r=diag\{I_2,-2\}=2\sigma_z^{(02)}-\sigma_z^{(01)}$. We choose
\begin{eqnarray}                                                                                     \label{A5}
\mathrm{s}(\mathrm{u}(2) \oplus
 \mathrm{u}(1))=span\{i(\sigma_x^{(01)},\sigma_y^{(01)},\sigma_z^{(01)},r)\},
 \end{eqnarray}
\begin{eqnarray}                                                                                     \label{A6}
\mathrm{s}(\mathrm{u}(2) \oplus
 \mathrm{u}(1))^{\bot}=span\{i(I_3,\sigma_x^{(02)},\sigma_y^{(02)},\sigma_x^{(12)},\sigma_y^{(12)})\},\nonumber \\
 \end{eqnarray}
and its Cartan subalgebra
\begin{eqnarray}                                                                                     \label{A7}
\mathfrak{a}=span\{i(I_3,i\sigma_{y}^{(02)}\}.
\end{eqnarray}
So the one-qutrit matrix can be decomposed as
\begin{eqnarray}                                                                                     \label{A8}
M&=&e^{i
\alpha}\tilde{M}_1^{(01)}R_z^{(01)}(-\theta)R_z^{(02)}(2\theta)R_y^{(02)}(\beta)\nonumber\\&&R_z^{(02)}(2\theta')
R_z^{(01)}(-\theta')\tilde{M}_2^{(01)}\nonumber\\
&=&e^{i \alpha}M_1^{(01)}M^{(02)}M_2^{(01)}.
\end{eqnarray}
Lie subalgebra and Cartan subalgebra of the Cartan decomposition can
be different, so the decomposition is not unique, we can get more
generic Eq. (\ref{5}) in Sec.\uppercase\expandafter{\romannumeral3}.

\section{CARTAN DECOMPOSITON OF ONE-QUDIT GATES}

The one-qudit gates form a  $\mathrm{U}(d)$ group. We can also use
the qusi-spin basis. There are  $\frac{1}{2}d(d-1)$
$\sigma_x^{(ij)}$ matrices, $\frac{1}{2}d(d-1)$ $\sigma_y^{(ij)}$
matrices and $d-1$ independent $\sigma_z^{(ij)}$ matrices in the
$\frac{1}{2}d(d-1)$ of them for a n-dimensional Hilbert space.
Multiplying these $d^2-1$ independent qusi-spin matrices by $i$, we
gain the basis vectors of Lie algebra $\mathrm{su}(d)$. Together
with $d\times d$ identity matrix multiplied by $i$, they constitute
the basis vectors of Lie algebra  $\mathrm{u}(d)$. We also take a
kind of \textbf{AIII} type Cartan decomposition for $\mathrm{u}(d)$,
that is
\begin{equation}                                         \label{B1}
\mathrm{u}(d)=\mathrm{s}(\mathrm{u}(d-1)\oplus
\mathrm{u}(1))+\mathrm{s}(\mathrm{u}(d-1)\oplus u(1))^\perp.
\end{equation}
Lie algebra $\mathrm{s}(\mathrm{u}(d-1)\oplus \mathrm{u}(1))$
consists of subagebra $\mathrm{su}(d-1)$ and a complex basis
$r=diag\{I_{d-1},-(d-1)\}$. We choose its Cartan subalgebra
\begin{eqnarray}                                                                                     \label{B2}
\mathfrak{\alpha}=span\{i(I_d, \sigma_{y}^{(d-2,d-1)})\}.
\end{eqnarray}
So arbitrary one-qudit matrix can be expressed as
\begin{eqnarray}                                                                                     \label{B3}
M=e^{i \alpha}K_{1}R_y^{(d-2,d-1)}(\beta)K_2,
\end{eqnarray}
where $K_i\in \mathrm{S}(\mathrm{U}(d-1)\oplus\mathrm{U}(1))$ group.
The matrix $M$ can be re-expressed as
\begin{eqnarray}                                                                                     \label{B4}
M&=&e^{i \mathfrak{\alpha}}\tilde{K'_1}e^{i\theta
r}R_y^{(d-2,d-1)}(\beta)e^{i\theta'r} \tilde{K'_2}\nonumber\\
&=&e^{i \mathfrak{\alpha}}K'_{1}M^{(d-2,d-1)} K'_2.
\end{eqnarray}
where $\tilde{K'_i},K'_i\in \mathrm{SU}(d-1)\oplus 1$. That is,
 $r$ can be expressed as a linear combination of
$\sigma_z^{(jk)}$s,
$r=\sigma_z^{(0,d-2)}+\cdots+\sigma_z^{(d-3,d-2)}+(d-1)\sigma_z^{(d-2,d-1)}$,
so   $e^{i\theta r}$ is a product of a serious of $R_z^{(jk)}$s.
$R_y^{(d-2,d-1)}$ combines with $R_z^{(d-2,d-1)}$ in $e^{i\theta r}$
and $e^{i\theta r'}$ to form  $M^{(d-2,d-1)}$, other $R_z^{(jk)}$s
are absorbed in  $K_i'$s.

From Eq. (\ref{B4}), we can see that the $d$-dimensional one-qudit
elementary gates needs one pair of basic gate more than that for the
($d$-1)-dimensional qudit. They come from Euler decompositions of
$M^{(d-2,d-1)}$. The ($d$-1)-dimensional qudit matrix $K'$ can be
decomposed further in same mode. The successive decomposition can be
done until the qutrit occurs. The one-qutrit elementary gates are
two pairs of basic gates, so we can infer that the set of
$d$-dimensional one-qudit elementary gates has $d-1$ pairs of basic
gates. Likewise the mode of Cartan decomposition and its Cartan
subalgebra can be different, the decomposition is not unique, so the
choice of $d-1$ pairs of basic gates is not unique too.

\end{document}